\documentclass[a4paper,11pt]{article}
\pdfoutput=1 

\usepackage{jheppub} 

\usepackage[T1]{fontenc} 

\title{\boldmath Higgs inflation and its extensions and the further refining dS swampland conjecture}

\author{Yang Liu}

\affiliation[a]{School of Physics and Astronomy, University of Nottingham, Nottingham NG7 2RD, UK}
\affiliation[b]{Nottingham Centre of Gravity, University of Nottingham, Nottingham NG7 2RD, UK}

\emailAdd{yang.liu@nottingham.ac.uk}

\usepackage{latexsym}

\abstract{On the one hand, in ref.[1], David Andriot and Christoph Roupec proposed an alternative refined de Sitter conjecture, which gives a natural condition on a combination of the first and second derivatives of the scalar potential [1]. On the other hand, in our previous article [2], we have found that Palatini Higgs inflation model is in strong tension with the refined de Sitter swampland conjecture [2]. Therefore, following our previous research, in this article we examine if Higgs inflation model and its two variations: Palatini Higgs inflation and Higgs-Dilaton model [3] can satisfy the “further refining de Sitter swampland conjecture” or not. Based on observational data [4,5,6], we find that these three inflationary models can always satisfy this new swampland conjecture if only we adjust the relevant parameters $a$, $b = 1-a$ and $q$. Therefore, if the “further refining de Sitter swampland conjecture” does indeed hold, then the three inflationary models might all be in “landscape”.}

\begin{document} 
\maketitle
\flushbottom

\section{Introduction}
\label{sec:intro}

The swampland program is a very interesting development for the phenomenology of quantum gravity theories, since it connects to bottom-up approaches, and to questions regarding effective field theories and their UV completions [1]. In recent years, papers have proposed the “refined de Sitter swampland conjecture” criteria considering the derivative of scalar field potentials, such as [7]. If we consider a four-dimensional (4d) theory of real scalar field $\phi^i$ coupled to gravity, whose dynamics is governed by a scalar potential $V(\phi^j)$, then the action can be written as [1]
\begin{equation}\label{eq:1.1}
S = \int_{4} \sqrt{|g_4|} \left(\frac{M^2_p}{2} R_4 - \frac{1}{2} h_{ij} \partial_{\mu} \phi^i \partial_{\mu} \phi^j - V \right),
\end{equation}
where $M_p$ is the Planck mass, $|g_4|$ is the determinant of the metric matrix of $4d$ spacetime, $R_4$ is the $4d$ Riemann curvature of spacetime and $h_{ij}$ is the metric on the target space of the scalar fields. \\
The refined dS swampland conjecture states that an effective theory of quantum gravity, i.e., not in the swampland, should satisfy one of the following two conditions [1,7]:
\begin{equation}\label{eq:1.2}
|\nabla V| \geq \frac{c_1}{M_p} \cdot V,
\end{equation}
or
\begin{equation}\label{eq:1.3}
min(\nabla_i \nabla_j V) \leq -\frac{c_2}{M^2_p} \cdot V,
\end{equation}
where $c_1$ and $c_2$ are both positive constants of the order of 1. $(1.2)$ corresponds to original “swampland conjecture”. If we introduce two “slow-roll” field inflation indices [1,2]:
\begin{equation}\label{eq:1.4}
\epsilon_V = \frac{1}{2} \left(\frac{V'}{V} \right)^2,
\end{equation}
\begin{equation}\label{eq:1.5}
\eta_V =  \frac{V''}{V}, 
\end{equation}
where $V$ is the potential energy in the action $(1.1)$, then the refined swampland conjecture $(1.2)$ and $(1.3)$ can be rewritten for $V > 0$ as [1]:
\begin{equation}\label{eq:1.6}
\sqrt{2 \epsilon_V} \geq c_1 \qquad or \qquad \eta_V \leq -c_2 . 
\end{equation}
However, the formulation of the refined swampland conjecture is peculiar: it is given by two distinct conditions $(1.2)$ and $(1.3)$ on two different quantities $\epsilon_V$ and $\eta_V$ [1]. One of the reasons for this formulation is the derivation of this conjecture proposed in ref.[1,7], where condition $(1.2)$ is derived in a weak coupling, semi-classical regime, while condition $(1.3)$ is defined by the requirement $\eta_V \geq -1$. Therefore, there exists a separation between the two conditions and quantities. As a result, this formulation cannot provide any information on both quantities simultaneously [1].\\
Due to the above discussion, David Andriot and Christoph Roupec proposed a further refining de Sitter swampland conjecture which suggested that a low energy effective theory of a quantum gravity that takes the form $(1.1)$ should verify, at any point in field space where $V > 0$ [1], 
\begin{equation}\label{eq:1.7}
\left(M_p \frac{|\nabla V|}{V} \right)^q - a M^2_p \frac{min(\nabla_i \nabla_j V)}{V} \geq b \qquad with \quad a+b=1, \quad a,b>0 \quad q>2.
\end{equation}
In terms of the slow-roll parameters, the conjecture can be rewritten as [1]:
\begin{equation}\label{eq:1.8}
(2\epsilon_V)^{\frac{q}{2}} - a \eta_V \geq b.
\end{equation}
Now we will turn our interest to inflation of universe. Inflation is a well-established paradigm which is able to explain the flatness, homogeneity and isotropy of the universe and the generation of the primordial density fluctuations seeding structure formation [3]. Inflation includes a period of vacuum-dominated accelerating expansion in the early universe, and eventually ceases in a period of reheating and transition to a radiation-dominated hot big bang cosmology. The vacuum dynamics can be generally modeled using one or more scalar order parameters $\phi$ such that the potential energy of the field dominates over the kinetic energy [3].\\
In this article we mainly focus on Higgs inflation model and its extensions to examine if these models satisfy the further refining swampland conjecture or not. On the one hand, Higgs inflation model owns important phenomenological meaning. Higgs field has been discovered in nature, and is proposed to be the only scalar field in the Standard Model [3,8]. The Higgs inflation interprets the Higgs as the inflaton [8], where the Higgs field has a nonminimal coupling to gravity [3,8]. This inflation scenario can explain the dark energy problem and allow primordial black hole production (PBH) [8]. On the other hand, in our previous research, we have found that one of the extensions of Higgs inflation, Palatini Higgs inflation, is in strong tension with the refined de Sitter swampland conjecture [2]. Considering the important theoretical meaning of swampland conjecture and phenomenological meaning of Higgs inflation model, we intend to find a new swampland conjecture which Higgs inflation model and its extensions can satisfy. Several variations and extentions of Higgs inflation model have been suggested [9,10,11]. However, in this article, we restrict ourselves to those proposals that are more closely related to the minimalistic spirit of the original scenario [3]. In particular, we consider Palatini Higgs inflation and Higgs-Dilaton model.\\
The article is composed as follows: in section 2, we briefly review Higgs inflation model and its two extensions. In section 3, we examine if these inflation models can satisfy the further refining dS swampland conjecture. In section 4, the results we have obtained are discussed. 

\section{Higgs inflation and its extensions}
In section 2, we briefly review the Higgs inflation model and its two extensions: Palatini Higgs inflation and Higgs-Dilaton model [3]. 

\subsection{Higgs inflation model}
The total Higgs inflation action [3,12]
\begin{equation}\label{eq:2.1}
S = \int d^4 x \sqrt{-g} [\frac{M^2_p}{2}R + \xi H^{\dagger} H R + L_{SM}],
\end{equation}
contains two dimensionfull parameters: the reduced Planck mass $M_p = 2.435 \times 10^{18}GeV$ and the Higgs mass expectation value $v_{EW} \approx 250GeV$ responsible for the masses within the SM Lagrangian density $L_{SM}$ [3]. At the large field value, the Planck mass plays an important role for inflation [3,8]. In the unitary gauge, we can write Higgs field as $H= (0, h)^T/\sqrt{2}$ [3]. Then the action $(2.1)$ can be rewritten as 
\begin{equation}\label{eq:2.2}
S = \int d^4 x \sqrt{-g} [\frac{M^2_p + \xi h^2}{2}R - \frac{1}{2} (\partial h)^2 - U(h)],
\end{equation} 
with
\begin{equation}\label{eq:2.3}
U(h) = \frac{\lambda}{4} (h^2 - v^2_{EW})^2,
\end{equation}
the symmetry breaking potential in the Standard Model [3,8]. \\
It is convenient to reformulate action $(2.2)$ in the Einstein frame by a Weyl transformation $g_{\mu\nu} \rightarrow \Theta g_{\mu\nu}$ with [8]:
\begin{equation}\label{eq:2.4}
\Theta^{-1} = 1 + \frac{h^2}{F^2_{\infty}}, \qquad F_{\infty} \equiv \frac{M_p}{\sqrt{\xi}}.
\end{equation}
Then in the Einstein frame, action $(2.2)$ can be rewritten as [3,8]:
\begin{equation}\label{eq:2.5}
S = \int d^4 x \sqrt{-g} [\frac{M^2_p}{2}R - \frac{1}{2} M^2_p K(\Theta) (\partial \Theta)^2 - V(\Theta)],
\end{equation}
which contains a non-exactly flat potential [3]:
\begin{equation}\label{eq:2.6}
V(\Theta) \equiv U(\Theta) \Theta^2 = \frac{\lambda F^4_{\infty}}{4} [1- \left(1+ \frac{v^2_{EW}}{F^2_{\infty}} \right) \Theta]^2,
\end{equation} 
and a non-canonical kinetic sector [3]
\begin{equation}\label{eq:2.7}
K(\Theta) \equiv \frac{1}{4 |a| \Theta^2} \left(\frac{1-6|a|\Theta}{1-\Theta} \right), 
\end{equation}
where
\begin{equation}\label{eq:2.8}
a \equiv -\frac{\xi}{1+6\xi}. 
\end{equation} 

\subsection{Palatini Higgs inflation}
In Higgs inflation model, the action is minimized with respect to the metric. This procedure implicitly assumes the existence of a Levi-Civita
connection depending on the metric tensor and the inclusion of a York-Hawking-Gibbons term ensuring the cancellation of a total derivative term with no-vanishing variation at the boundary [3,13,14]. One could alternatively consider a Palatini formulation of gravity in which the metric tensor and the connection are treated independently and no additional boundary term is required [3,15]. To see this explicitly, we consider the action $(2.2)$ with $R=g^{\mu\nu}R_{\mu\nu}(\Gamma, \partial\Gamma)$ and $\Gamma$ a non-Levi-Civita connection. Performing a Weyl transformation $g_{\mu\nu} \rightarrow \Theta g_{\mu\nu}$ in the Einstein frame and a field redefinition [3,15], then at $\phi \gg v_{EW}$, the action can be rewritten as [3]
\begin{equation}\label{eq:2.9}
S = \int d^4 x \sqrt{-g} [\frac{M^2_p}{2}R - \frac{1}{2} (\partial \phi)^2 - V(\phi)], 
\end{equation} 
with
\begin{equation}\label{eq:2.10}
V(\phi) = \frac{\lambda}{4} F^4(\phi), \qquad F(\phi) \equiv F_{\infty} \tanh \left(\frac{\sqrt{a}\phi}{M_p} \right) 
\end{equation}
More details can be found in ref.[3].

\subsection{Higgs-Dilaton model}
The existence of robust predictions in (non-critical) Higgs inflation is intimately related to the emerging dilatation symmetry of its tree-level action at large field values [3]. The uplifting of Higgs inflation to a completely scale-invariant setting was considered in several articles [16,17,18,19,20]. In the unitary gauge $H= (0, h)^T/\sqrt{2}$, the action of the graviscalar sector of the Higgs-dilaton model takes the form [3]:
\begin{equation}\label{eq:2.11}
S = \int d^4 x \sqrt{-g} [\frac{\xi_h h^2 + \xi_{\chi} \chi^2}{2}R - \frac{1}{2} (\partial h)^2 - \frac{1}{2} (\partial \chi)^2 - V(h, \chi)], 
\end{equation} 
with
\begin{equation}\label{eq:2.12}
U(h,\chi)= \frac{\lambda}{4} (h^2 - \alpha \chi^2)^2 + \beta \chi^4 
\end{equation}
a scale-invariant version of the Standard Model symmetry breaking potential and $\alpha$, $\beta$ positive dimensionless parameters [3,17,18,19,20].\\
Performing a Weyl rescaling $g_{\mu\nu} \rightarrow M^2_p/(\xi_h h^2 + \xi_{\chi} \chi^2)g_{\mu\nu}$ and a field redefinition, then the action in the Einstein frame can be obtained [3]
\begin{equation}\label{eq:2.13}
S = \int d^4 x \sqrt{-g} [\frac{M^2_p}{2}R - \frac{1}{2} M^2_p K(\Theta) (\partial \Theta)^2 - \frac{1}{2} \Theta (\partial \Phi)^2 - U( \Theta)]. 
\end{equation}
It contains a potential 
\begin{equation}\label{eq:2.14}
U(\Theta) = U_0 (1-\Theta)^2, \qquad U_0 \equiv \frac{\lambda M^4_p}{4} \left(\frac{1 + 6\bar{a}}{\bar{a}} \right)^2, 
\end{equation}
where
\begin{equation}\label{eq:2.15}
a \equiv - \frac{\xi_h}{1+6\xi_h}, \qquad \bar{a} \equiv a \left(1- \frac{\xi_h}{\xi_{\chi}} \right), 
\end{equation}
and a kinetic sector for $\Theta$ field 
\begin{equation}\label{eq:2.16}
K(\Theta) = \frac{1}{4 |\bar{a}| \Theta^2} \left(\frac{c}{|\bar{a}|\Theta - c} + \frac{1-6|\bar{a}|\Theta}{1-\Theta} \right),
\end{equation}
which contains two “inflationary” poles at $\Theta =0$ and $\Theta = c/|\bar{a}|$ and a “Minkowski” pole at $\Theta =1$ [3]. The “Minkowski” pole does not play a significant role during inflation and can be neglected for all practical purposes [3].

\section{Examine for the further refining dS swampland conjecture}
In this section, we intend to examine if Higgs inflation model and its two extensions satisfy the further refining dS swampland conjecture. In this section, we will take $M_p=1$.

\subsection{Higgs inflation model}
For the scalar potential $V(\phi)$, we can define two parameters
\begin{equation}\label{eq:3.1}
F_1= \frac{|dV(\phi)/d\phi|}{V},
\end{equation}
and
\begin{equation}\label{eq:3.2}
F_2= \frac{d^2V(\phi)/d\phi^2}{V},
\end{equation}
Considering eqs.$(1.4)$ and $(1.5)$, we have
\begin{equation}\label{eq:3.3}
F_1= \sqrt{2\epsilon_V}, \qquad F_2 = \eta_V.
\end{equation}
In the slow-roll region, $F_1$ and $F_2$ can be written in terms of the spectral index of the primordial curvature power spectrum $n_s$ and tensor-tensor-ratio $r$, namely [8],
\begin{equation}\label{eq:3.4}
F_1= \sqrt{2\epsilon_V} = \sqrt{\frac{r}{8}},
\end{equation} 
\begin{equation}\label{eq:3.5}
F_2= \eta_V = \frac{n_s-1+\frac{3}{8}r}{2}.
\end{equation}
At the tree level, the Higgs inflation model predicts that [8]
\begin{equation}\label{eq:3.6}
n_s \simeq 0.965, \qquad and \qquad r \simeq 0.003,
\end{equation}
which are consistent with the observational data [8]
\begin{equation}\label{eq:3.7}
n_s \simeq 0.965 \pm 0.004, \qquad and \qquad r \lesssim 0.06.
\end{equation}
Inserting $(3.6)$ into $(3.4)$ and $(3.5)$, we obtain that
\begin{equation}\label{eq:3.8}
F_1= 0.0194, \qquad F_2 = -0.0169.
\end{equation}
Considering the refined dS swampland conjecture $(1.6)$, we have
\begin{equation}\label{eq:3.9}
c_1 \leq 0.0194, \qquad or \qquad c_2 \leq 0.0169.
\end{equation}
However, $c_1$ and $c_2$ are both not $O(1)$, which means that Higgs inflation model is in strong tension with the refined de Sitter swampland conjecture.\\
Then let us examine if the model satisfies the refining de Sitter swampland conjecture. According to eqs.$(1.7)$ and $(1.8)$, we have
\begin{equation}\label{eq:3.10}
(2\epsilon_V)^{\frac{q}{2}} - a \eta_V \geq 1-a, \qquad q>2.
\end{equation}
Inserting $(3.8)$ into $(3.10)$, we have
\begin{equation}\label{eq:3.11}
0.0194^q + 0.0169a \geq 1-a,
\end{equation}
namely, if only $a$ could satisfy the condition 
\begin{equation}\label{eq:3.12}
\frac{1}{1.0169} (1- 0.0194^q) \leq a < 1, \qquad q>2,
\end{equation}
then the further refining de Sitter conjecture can be satisfied.

\subsection{Palatini Higgs inflation}
In our previous research [2], we have found that if we use the recent inflationary data of Cosmic Microwave Background radiation [4,5,6], then the Palatini Higgs inflation model cannot satisfy the refined de Sitter swampland conjecture [2]. According to ref.[2], we have determined that
\begin{equation}\label{eq:3.13}
1.191 \times 10^{-6} < F_1 < 9.823 \times 10^{-6}, \qquad F_2 \simeq -0.019,
\end{equation}
then we have found the upper bound of $c_1$ and $c_2$ are 
\begin{equation}\label{eq:3.14}
c_1 < 9.823 \times 10^{-6}, \qquad c_2 \simeq 0.019,
\end{equation}
However, neither $c_1$ nor $c_2$ is positive constant of the order of 1 [2], therefore, the Palatini Higgs inflation model is in strong tension with the refined de Sitter swampland conjecture [2]. Then in this article, we intend to explore if this model can satisfy the further refining dS swampland conjecture.\\
Combining eqs.$(3.4)$, $(3.5)$ and $(3.10)$, then we have
\begin{equation}\label{eq:3.15}
F^q_1 -aF_2 \geq 1-a.
\end{equation}
If we take the upper bound of $F_1 =9.823 \times 10^{-6}$ and $F_2 =-0.019$, then we obtain that
\begin{equation}\label{eq:3.16}
(9.823 \times 10^{-6})^q \geq 1-1.019a.
\end{equation}
When $a=0.981$, $1-1.019a=0$. We can examine that when $a < 0.981$, we can always find a $q$ whose value is larger than 2.\\
Similarly, if we take the lower bound of $F_1 =1.191 \times 10^{-6}$ and $F_2 =-0.019$, then we obtain that
\begin{equation}\label{eq:3.17}
(1.191 \times 10^{-6})^q \geq 1-1.019a.
\end{equation}
We can also examine that when $a < 0.981$, we can always find a $q$ whose value is larger than 2.\\
Therefore, we can conclude that if we use the recent inflationary data of Cosmic Microwave Background radiation [4,5,6], we can always find the approriate values of $a$ and $q$ so that the Palatini inflation model can satisfy the refining dS swampland conjecture.

\subsection{Higgs-Dilaton model}
Now we will turn our interest to Higgs-Dilaton model. From $(2.14)$ we have
\begin{equation}\label{eq:3.18}
U'(\Theta) = 2U_0\Theta-2U_0, 
\end{equation}
\begin{equation}\label{eq:3.19}
U''(\Theta) = 2U_0 >0. 
\end{equation}
In this article, we only consider the case of $0 \leq \Theta < 1$. In particular, the two inflationary poles $\Theta = 0$ and $\Theta = \frac{c}{|\bar{a}|}$. Since $U(\Theta) >0$ and $U''(\Theta) >0$, the second criteria of the refined dS swampand conjecture cannot be satisfied.\\
Firstly, we will examine if the Higgs-Dilaton model satisfies the original dS swampland conjecture or the first criteria of the refined dS swampland conjecture.  \\
At $\Theta =0$, $U(0)=U_0$, $|U'(0)|=2U_0$, then $|U'(0)|=2U(0)$. 2 is a number of order of 1, therefore Higgs-Dilaton model satisfies the original dS swampland conjecture or the first criteria of the refined dS swampland conjecture at $\Theta =0$.\\
At $\Theta = c/|\bar{a}|$, $U(c/|\bar{a}|) = U_0 \left( 1- \frac{c}{|\bar{a}|} \right)^2$, $U'(c/|\bar{a}|) = 2U_0 \left( \frac{c}{|\bar{a}|}-1\right)$. In order to satisfy the original dS swampland conjecture, we need $|U'(c/|\bar{a})| \geq c_1 U(c/|\bar{a})$, namely,
\begin{equation}\label{eq:3.20}
c_1 \left( 1- \frac{c}{|\bar{a}|} \right) \leq 2.
\end{equation}
To ensure that $c_1$ is a parameter of the order of 1, then we obtain $0 < c/ |\bar{a}| < \frac{4}{5}$. Therefore, if only $0 \leq \Theta < \frac{4}{5}$, the Higgs-Dilaton model could satisfy the original dS swampland conjecture or the refined dS swampland conjecture. \\
Secondly, we will examine if the Higgs-Dilaton model satisfies the further refining dS swampland conjecture. \\
Combining eqs.$(1.7)$, $(1.8)$, $(3.17)$ and $(3.18)$, we obtain that if the model satisfies the further refining dS swampland conjecture, then it must obey the following inequality
\begin{equation}\label{eq:3.21}
\left(\frac{2}{1-\Theta}\right)^q - \frac{2a}{(1-\Theta)^2} \geq 1-a \geq 0.
\end{equation}
For $q>2$, $(1-\Theta)^q < (1-\Theta)^2$, therefore for $q>2$ and $a<1$, the inequality 
\begin{equation}\label{eq:3.22}
\left(\frac{2}{1-\Theta}\right)^q - \frac{2a}{(1-\Theta)^2} > 0.
\end{equation}  
can satisfy naturally. Therefore, for the Higgs-Dilaton model, we can always find appropriate $a$ and $q$ to satisfy the further refining dS swampland conjecture. 

\section{Conclusions and discussions}
In ref.[1], the authors proposed a conjecture that a low energy effective theory of a quantum gravity that takes the form $(1.1)$ should satisfy the following relation at any point in field space where $V>0$ [1],
\begin{equation}\label{eq:4.1}
\left(M_p \frac{|\nabla V|}{V} \right)^q - a M^2_p \frac{min(\nabla_i \nabla_j V)}{V} \geq b \qquad with \quad a+b=1, \quad a,b>0 \quad q>2.
\end{equation}
In terms of the slow-roll parameters, the conjecture can be rewritten as [1]:
\begin{equation}\label{eq:4.2}
(2\epsilon_V)^{\frac{q}{2}} - a \eta_V \geq b.
\end{equation}
In our previous work [2], we have found that Palatini inflation model is in strong tension with the refined de Sitter swampland conjecture [2]. However, Higgs inflation model has important phenomenological meaning [8]. Therefore, we intend to explore if Higgs inflation model and its two variations: Palatini Higgs inflation and Higgs-Dilaton model can satisfy the further refining swampland conjecture or not. \\
Based on recent observational data [4,5,6], we find that these three inflationary models can always satisfy this new swampland conjecture if only we adjust the values of three relevant parameters $a$, $b = 1-a$ and $q$. Therefore, we suggest that if the “further refining de Sitter swampland conjecture” indeed holds, then the three inflationary models might all be in “landscape”.\\
In fact, according to our analysis, the refining de Sitter swampland conjecture is too “loose” for the three inflationary models. In other words, we cannot determine the upper and lower bounds of the three parameters $a$, $b=1-a$ and $q$ using this new swampland conjecture. It is thus natural to explore other swampland conjecture in string theory to constrain the range of these physical parameters in the future work.


\end{document}